\documentclass[journal=nalefd,manuscript=letter]{achemso}

\usepackage{multicol}

\usepackage{xcolor}
\usepackage[version=3]{mhchem} 

\author{Tomás Santiago-Cruz}
\affiliation[MPL]
{Max Planck Institute for the Science of Light, Staudtstraße 2, 91058 Erlangen, Germany.}
\alsoaffiliation[FAU]
{University of Erlangen-Nürnberg, Staudtstraße 7/B2, 91058 Erlangen, Germany.}
\email{jose-tomas.santiago@mpl.mpg.de}
\author{Vitaliy Sultanov}
\affiliation[MPL]
{Max Planck Institute for the Science of Light, Staudtstraße 2, 91058 Erlangen, Germany.}
\alsoaffiliation[FAU]
{University of Erlangen-Nürnberg, Staudtstraße 7/B2, 91058 Erlangen, Germany.}
\author{Haizhong Zhang}
\affiliation[AStar]
{Institute of Materials Research and Engineering, A*STAR (Agency for Science, Technology and Research) Research Entities, 2 Fusionopolis Way, \#08-03 Innovis, Singapore 138634.}
\author{Leonid A. Krivitsky}
\affiliation[AStar]
{Institute of Materials Research and Engineering, A*STAR (Agency for Science, Technology and Research) Research Entities, 2 Fusionopolis Way, \#08-03 Innovis, Singapore 138634.}
\author{Maria V. Chekhova}
\affiliation[MPL]
{Max Planck Institute for the Science of Light, Staudtstraße 2, 91058 Erlangen, Germany.}
\alsoaffiliation[FAU]
{University of Erlangen-Nürnberg, Staudtstraße 7/B2, 91058 Erlangen, Germany.}
\alsoaffiliation[Moscow]
{Department of Physics, M. V. Lomonosov Moscow State University, Leninskie Gory, 119991 Moscow, Russia.}

\title[An \textsf{achemso} demo]
  {Spontaneous Parametric Down-Conversion from Subwavelength Nonlinear Films}

\abbreviations{IR,NMR,UV,FP}

\begin{document}

\begin{abstract}
Miniaturised entangled photon sources are highly demanded for  the development of integrated quantum photonics. Since the invention of subwavelength optical metasurfaces and their successes at replacing bulky optical components, the  possibility of implementing entangled photon sources on such devices is actively investigated. Here, as a first step towards the development of quantum optical metasurfaces (QOM), we demonstrate photon pair generation via spontaneous parametric down-conversion (SPDC) from subwavelength films. We achieved photon pair generation with a high coincidence-to-accidental ratio in lithium niobate and gallium phosphide nanofilms. In addition, we have measured the SPDC frequency spectrum via fibre spectroscopy, obtaining photon pairs with a spectral bandwidth of 500\;nm, limited only by the overall detection efficiency. Moreover, we have observed the vacuum field enhancement due to a Fabry-Perot resonance inside the nonlinear films. Our experiments lay the groundwork for the future development of flat SPDC sources, including QOM.

\end{abstract}



\section{Introduction}

In quantum optics, entangled photon sources have been of paramount importance for the realisation of different experiments, ranging from basic science to practical applications. Among them, spontaneous parametric down-conversion (SPDC) is perhaps the most popular source of entangled photons. In SPDC, a high-energy pump photon interacting with a second-order nonlinear material is down-converted into two lower-energy daughter photons, where the three interacting photons fulfil the energy conservation and phase-matching (momentum conservation) conditions. Similar to the spontaneous emission of atoms or solid-state emitters, SPDC is stimulated by zero-point vacuum fluctuations \cite{klyshkoNonlinearoptics}. SPDC utilises bulk nonlinear materials to achieve a relatively high conversion efficiency, which in turn complicates its scalability, and makes hard its integration into photonic circuits \cite{Wang2020}.
While reducing the interaction length compromises the conversion efficiency, it also brings some advantages. First of all, reducing the interaction length relaxes the phase-matching condition \cite{Okoth2019}, allowing the utilization of any second-order nonlinear material. Second, photon pairs from thin materials display ultra-broad frequency spectrum, ultra-short correlation time and high degree of continuous-variable entanglement \cite{Okoth2019,Okoth2020}.

There has been a tremendous effort aimed at scaling  the length of SPDC sources down to the nanometer level without sacrificing the conversion efficiency. In view of the promising performance of subwavelength optical metasurfaces in classical nonlinear frequency conversion experiments \cite{Liu2016,Vabishchevich2018,Liu2018,Koshelev2020},
many works suggested their use as a viable option to boost the efficiency of SPDC by exploiting Mie-type resonances \cite{Poddubny2018,Marino2019,Ma2020,Solntsev2020}  and bound states in the continuum (BIC) \cite{Koshelev2020a,Parry2019,Wang2019}. Up to date, there is no convincing evidence of photon pair generation via SPDC from subwavelength metasurfaces  and even subwavelength films \cite{Faccio2018}, although photon pairs  have been obtained in carbon nanotubes via spontaneous four-wave mixing \cite{Lee2017}.

Here, we bridge the gap between nonlinear quantum optics and flat optics by achieving, for the first time, the generation of entangled photons via SPDC from subwavelength  films. Through the fibre spectroscopy of SPDC photons, we probe their ultra-broad spectrum and observe resonances due to Fabry-Perot field buildup. This is the first step towards the spectroscopy of SPDC in metasurfaces. 


\section{Results and discussion}

In the first experiment (see Correlation experiment in Methods), we  demonstrate photon pair generation via degenerate type-0 SPDC by registering photon coincidences between two detectors. As nonlinear media, we have tested two different subwavelength films, a 300\,nm thick lithium niobate (LN) and a 400\,nm thick gallium phosphide (GaP) films. Detailed description of the films is given in Methods. We have chosen these nonlinear materials as they have been already adapted into metasurfaces \cite{Cambiasso2017,Shcherbakov2020,Fedotova2019,Ma2020}. 

\begin{figure}[t!]
     \includegraphics[width=0.4\textwidth]{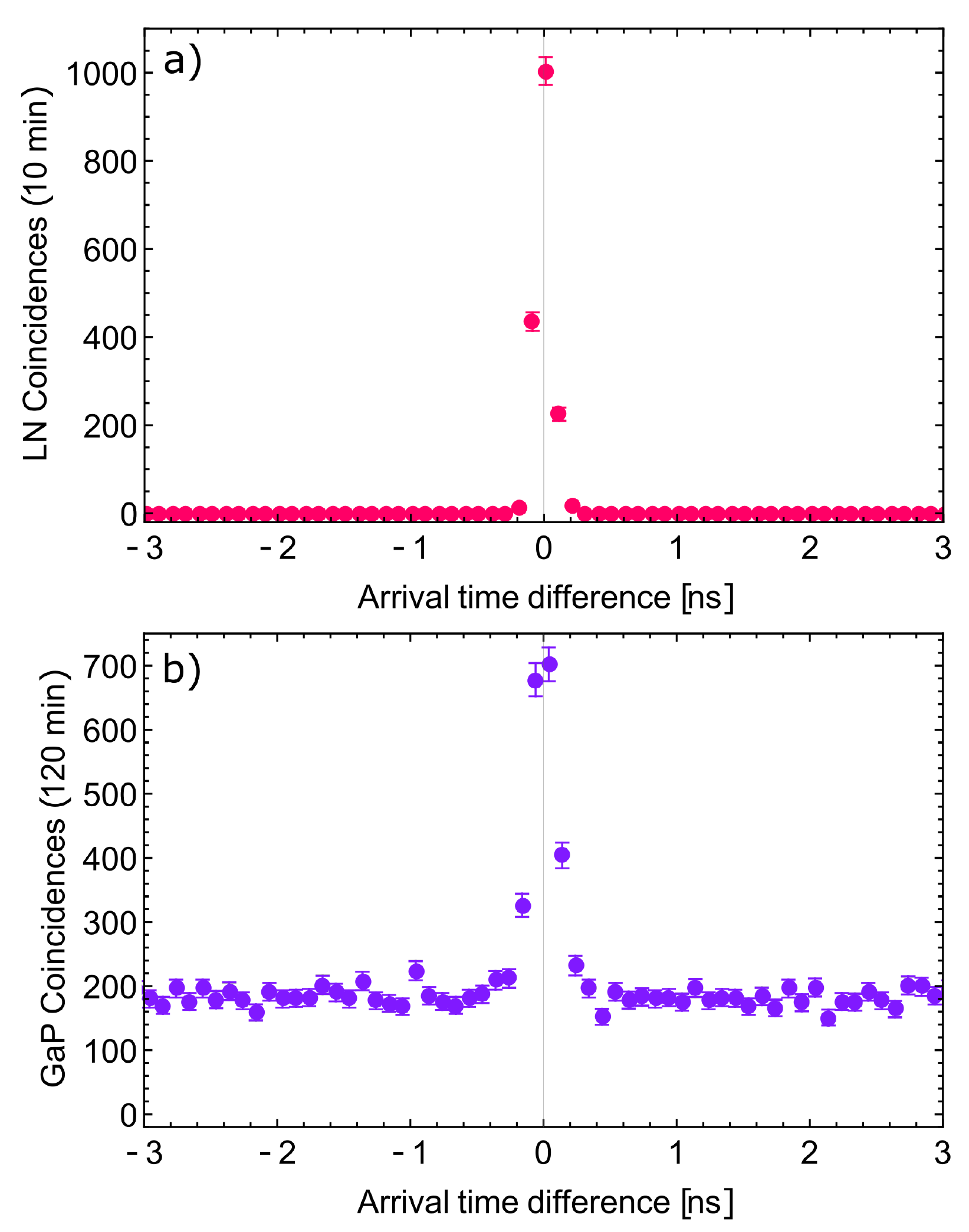}
  \caption{Coincidence histograms as measured in a) LN and  b) GaP, when pumped at $9 \; \mathrm{mW}$ after an acquisition time of 10 min and 120 min, respectively.}
  \label{fgr:LN-Gap_coincidences}
\end{figure}

Figure~\ref{fgr:LN-Gap_coincidences} shows the coincidence histograms, i.e., the numbers of two-photon detection events versus the difference in the photon detection times, acquired in LN for 10 min (a) and in GaP for 120 min (b). The pump was a continuous-wave laser with the wavelength 685\;nm and power $9 \; \mathrm{mW}$, and degenerate photon pairs were detected after a 50\;nm FWHM bandpass filter centred at 1375\;nm.

The peak at the center indicates simultaneous arrival of one photon at each detector and is a signature of photon pair generation. The height of the peak quantifies the coincidence rate, and it should considerably exceed the background (accidental coincidence rate) to witness SPDC. We emphasize that the mere observation of a coincidence peak at zero arrival time difference is not yet an indication of photon pair generation because the bunching of thermal light also gives a peak at zero arrival time difference \cite{Loudon2000}, and can be misinterpreted as the detection of photon pairs. But for thermal light, the peak-to-background ratio does not exceed a two. Therefore, a higher peak should be observed to witness photon pair generation. The width of the correlation curves in Figure~\ref{fgr:LN-Gap_coincidences} is given by the timing jitter of the detectors. 

The  pair-generation rate, defined as the total number of true coincidences minus the total number of accidental coincidences, is $2.8 \pm 0.1 \; \mathrm{Hz}$ in LN and $0.20 \pm 0.01 \; \mathrm{Hz}$ in GaP, as calculated from the data shown in Figure~\ref{fgr:LN-Gap_coincidences}. 
Although these rates might seem too low, we should consider that the nonlinear interaction length is very small. Since the pair-generation rate increases linearly with the pump power \cite{klyshkoNonlinearoptics}, by pumping with hundreds of mW we could easily obtain tens of Hz. In our experiments we were limited to $\sim 9 \; \mathrm{mW}$, the maximum power delivered by the laser. In GaP, the photon generation rate was additionally reduced because of non-optimal orientation of the film \cite{Anthur2020}.

A drawback of semiconductors such as GaP is the strong fluorescence emitted when the pump is near the absorption edge. Indeed, the high level of accidental coincidences in the case of GaP we attribute to uncorrelated photons coming mainly from fluorescence. Unlike GaP, in LN the accidental coincidences are very few. Fluorescence becomes a serious problem when reducing the nonlinear interaction length $L$. Since fluorescence is an incoherent process, it scales down as $L$, while the pair-generation rate into a narrow frequency band scales  as $L^2$ \cite{klyshkoNonlinearoptics}, and the former surpasses the latter when $L$ is too small. An advantage of working with photon pairs is that a high rate of fluorescence does not mask the real signal because fluorescence photons  contribute only to the accidental coincidences and not to the true coincidences. Thus, the detection of correlated photon pairs is still possible by acquiring for a long time and using low pump powers. 

\begin{figure}[t!]
\centering
     \includegraphics[scale=0.59]{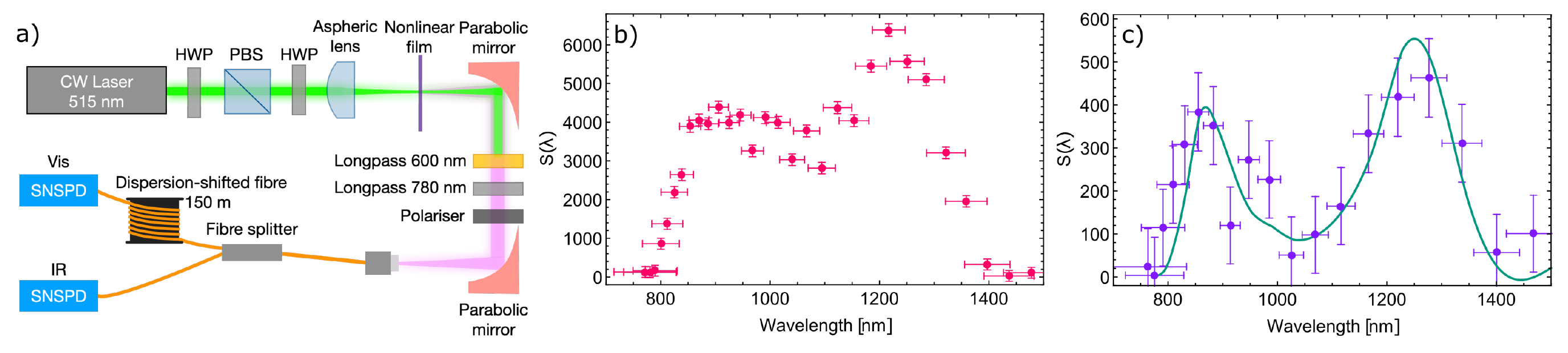}
  \caption{a) Fibre spectroscopy setup. SPDC spectrum as measured  in LN (b) and GaP (c) when pumped at $515$ nm with $25 \; \mathrm{mW}$ and $100 \; \mu  \mathrm{W}$, respectively.}
  \label{fgr:spectra}
\end{figure}

In the second experiment (see panel a) in Figure~\ref{fgr:spectra} and Methods), we measured the  SPDC spectrum via fibre spectroscopy \cite{Valencia2002}, where one photon of the pair is sent to a long dispersive fibre. Due to the group velocity dispersion (GVD) of the fibre, the biphoton wavepacket broadens in time but the spectrum remains unchanged, similarly to the propagation of short optical pulses in a fibre. As a result, the coincidence histogram will also broaden in time. When the GVD is monotonic and  has the same sign over the spectral region of interest, the relation between arrival time difference and wavelength is one-to-one, allowing the reconstruction of the spectrum. Since this technique relies on a large GVD, we changed the wavelength of the pump to  $515 \; \mathrm{nm}$ so as to have non-degenerate photon pairs below the zero-dispersion wavelength (ZDW) of the fibre, ZDW $\sim 1500 \; \mathrm{nm} $, which in turn allowed the measurement of a broader spectral bandwidth. 


Panels b) and c) in Figure~\ref{fgr:spectra} show the spectrum of SPDC generated in LN and GaP, respectively. The pump power was $25 \; \mathrm{mW}$ in LN and $100 \; \mu  \mathrm{W}$ in GaP. Note that the width of the measured spectrum is $\sim 500 \; \mathrm{nm} $, regardless of the nonlinear film, which is a remarkable result and highlights the most striking difference between SPDC sources based on ultra-thin films and bulk crystals.

Because of the high refractive index contrast between GaP and the substrate, the GaP film works as a Fabry-Perot (FP) etalon and there is a resonant field buildup at some frequencies, leading to the fringes in Figure~\ref{fgr:spectra}c. Importantly, here the resonant enhancement occurs for the vacuum fluctuations seeding SPDC, and not for an externally fed radiation. This effect has been studied theoretically~\cite{Kitaeva1982} but not observed before. For comparison, the green curve in Figure~\ref{fgr:spectra}c shows the theoretical dependence taking into account the resonant enhancement of the vacuum field and the relative efficiency of the detectors. The modulation in the measured spectrum, leading to about $5$ times enhancement of SPDC efficiency, is in good agreement with the theoretical dependence. Unlike GaP, in LN the measured spectrum is modulated  mainly by the efficiency of the detectors. 

Since the spectrum of SPDC from subwavelength films spans more than one octave \cite{Okoth2019}, the difference in the width of the spectrum in GaP and LN can be noticeable only if the entire spectral width is measured. In our experiments, the measured spectrum was restricted on the red side by the low GVD of the dispersive fibre near the ZDW, and on the blue side by the attenuation of the dispersive fibre and the splitting ratio of the fibre splitter. We could overcome the latter problem by splitting the photons around the degenerate wavelength ($1030 \; \mathrm{nm}$) using a dichroic mirror and sending the blue photon to the dispersive fibre, where the GVD is high, and then to the corresponding detector, while the conjugate photon is sent directly to the other detector. 

Unlike bulk SPDC sources, subwavelength SPDC sources are versatile and can be pumped at any wavelength: the ultra-small thickness of the films eliminates the need for phase-matching. In this work, we have pumped the same nonlinear films at 685\;nm and 515\;nm, but any other wavelength can be used at convenience.

While our study has focused only on LN and GaP films, it can be easily extended to other nonlinear materials that have been  adapted into metasurfaces, such as AlGaAs, GaAs and BaTiO$_3$, opening the door for the demonstration of SPDC in metasurfaces. 

We envisage that SPDC and nanoscale metasurfaces could benefit from each other. Metasurfaces can be used to generate SPDC and to boost the conversion efficiency by exploiting resonances of Mie-type and BIC. Meanwhile, the ultra broadband nature of SPDC could be a tool to reveal the resonances of metasurfaces. Here, we stress that, unlike harmonic generation (HG),  SPDC is spectrally broadband even with a narrowband pump. Thus, the observation of entangled photons over a broad spectral region does not require a tuneable pump, as we show in Figure~\ref{fgr:spectra}. Moreover, since we have demonstrated SPDC in materials used for metasurfaces, there is no need of integrating an external SPDC source into the metasurface \cite{Li2020}.

From the viewpoint of fundamental research, SPDC sources based on metasurfaces could be an interesting platform on which to investigate the interaction between vacuum fluctuations and optical resonances.

In summary, we have demonstrated ultra-broadband  photon pair generation via SPDC at the subwavelength scale, opening the door to a new field, namely flat nonlinear quantum optics. The subwavelength thickness of the sources pave the way to their integration into quantum photonic circuits and the utilisation of metasurfaces as a way to improve the efficiency, which could lead to a paradigm shift in quantum optics.


\section{Methods}

\subsection{Nonlinear films}

We have used a 300\;nm thick X-cut LN film on a fused silica substrate ($500 \; \mu \mathrm{m}$), with the optic axis in the plane of the film. The pump and daughter photons were polarized parallel to the optic axis such that the nonlinear process was mediated by the largest component $d_{33}$ of the nonlinear tensor in LN.  The optic axis was oriented along the vertical direction during the experiments. The second crystal under test was a 400\;nm thick GaP film on fused silica ($4 \; \mu \mathrm{m}$) on  sapphire substrate ($150 \; \mu \mathrm{m}$). Its orientation is described elsewhere \cite{Anthur2020}. In the case of GaP, through second harmonic generation experiments we found that the polarization configuration maximizing the efficiency is with both the pump and the photon pairs vertically polarized. This configuration, with the pump and daughter photons copolarized, is known as type-0 SPDC.

\subsection{Correlation experiment}

    We pumped the subwavelength nonlinear films with a continuous-wave (cw) pigtailed laser diode at $685 \; \mathrm{nm}$, delivering a maximum power of $\sim 9 \; \mathrm{mW}$. Degenerate photon pairs were detected by using a bandpass filter centred at $1375 \; \mathrm{nm} $ ($50 \; \mathrm{nm}$ FWHM). Further details of the setup are described elsewhere \cite{Okoth2019}. 

\subsection{Fibre spectroscopy of biphotons}

    For the efficient detection of broadband SPDC, we designed a chromatic aberration-free setup based on the correlation experiment described above. The setup is shown in Figure~\ref{fgr:spectra}a. We pumped the nonlinear films with a cw laser at  $515 \; \mathrm{mm}$. The pump power was adjusted by a half-wave plate (HWP) and a polarising beamsplitter (PBS). The pump polarisation was rotated by a HWP inserted after the PBS. The pump was focused into the nonlinear  films using an aspheric lens with  $4.51 \; \mathrm{mm}$ focal length. The SPDC radiation generated in the films was collected using a  $90 ^{\circ}$ off-axis gold-coated parabolic mirror with  $15 \; \mathrm{mm}$  reflective focal length. A hard-coated longpass filter with cut-on wavelength at $600 \; \mathrm{nm}$ filtered out the pump. In addition,  we inserted a longpass color glass filter with cut-on wavelength at $780 \; \mathrm{nm}$ in order to filter out unwanted incoherent radiation,  e.g. fluorescence,  generated in the films.  A broadband polarizer was used to detect photon pairs with vertical polarization. The polarizer also served as an additional filter removing the fluorescence. 
		
The SPDC radiation was fed into the input facet of a 1x2 broadband single-mode fibre splitter using the same parabolic mirror as the one used after the film.  A 150 m long dispersion-shifted fibre (DCF4, Thorlabs) was inserted in one output arm of the fibre splitter. Due to the long fibre, the coincidence histogram was broadened in time (see Supporting Information (SI)). The SPDC spectrum was retrieved from the coincidence histograms by converting arrival time differences into wavelengths using a calibration curve that was obtained by inserting longpass filters with redder cut-on wavelengths (see SI). The photons were detected by using one visible (Vis) and one infrared (IR) superconducting nanowire single-photon detector (SNSPD). Coincidence events between the two detectors were registered using a time-tagger device (not shown).
 
   Since the pump wavelength  was shifted down to $515 \; \mathrm{mm}$, non-degenerate photon pairs were generated below the zero-dispersion wavelength (ZDW) of the fibre, which occurs near $ 1500 \; \mathrm{nm}$. A longer-wavelength pump, e.g. $685 \; \mathrm{mm}$, produces non-degenerate photon pairs above and below the ZDW, and the technique to retrieve the spectrum fails because the correspondence between arrival time differences and wavelengths is not one-to-one.

\color{black}

\begin{acknowledgement}

T.S.C. is part of the Max Planck School of Photonics supported by BMBF, Max Planck Society, and Fraunhofer Society. We acknowledge the support of the Quantum Technologies for Engineering (QTE) program of A*STAR (Singapore). We thank Yuri Kivshar and  Aravind P. Anthur for fruitful discussions.

\end{acknowledgement}

\bibliography{achemso-demo}

\providecommand{\latin}[1]{#1}
\makeatletter
\providecommand{\doi}
  {\begingroup\let\do\@makeother\dospecials
  \catcode`\{=1 \catcode`\}=2 \doi@aux}
\providecommand{\doi@aux}[1]{\endgroup\texttt{#1}}
\makeatother
\providecommand*\mcitethebibliography{\thebibliography}
\csname @ifundefined\endcsname{endmcitethebibliography}
  {\let\endmcitethebibliography\endthebibliography}{}
\begin{mcitethebibliography}{26}
\providecommand*\natexlab[1]{#1}
\providecommand*\mciteSetBstSublistMode[1]{}
\providecommand*\mciteSetBstMaxWidthForm[2]{}
\providecommand*\mciteBstWouldAddEndPuncttrue
  {\def\EndOfBibitem{\unskip.}}
\providecommand*\mciteBstWouldAddEndPunctfalse
  {\let\EndOfBibitem\relax}
\providecommand*\mciteSetBstMidEndSepPunct[3]{}
\providecommand*\mciteSetBstSublistLabelBeginEnd[3]{}
\providecommand*\EndOfBibitem{}
\mciteSetBstSublistMode{f}
\mciteSetBstMaxWidthForm{subitem}{(\alph{mcitesubitemcount})}
\mciteSetBstSublistLabelBeginEnd
  {\mcitemaxwidthsubitemform\space}
  {\relax}
  {\relax}

\bibitem[Klyshko(2018)]{klyshkoNonlinearoptics}
Klyshko,~D.~N. \emph{Photons and Nonlinear Optics}; Routledge, 2018\relax
\mciteBstWouldAddEndPuncttrue
\mciteSetBstMidEndSepPunct{\mcitedefaultmidpunct}
{\mcitedefaultendpunct}{\mcitedefaultseppunct}\relax
\EndOfBibitem
\bibitem[Wang \latin{et~al.}(2020)Wang, Sciarrino, Laing, and
  Thompson]{Wang2020}
Wang,~J.; Sciarrino,~F.; Laing,~A.; Thompson,~M.~G. Integrated photonic quantum
  technologies. \emph{Nature Photonics} \textbf{2020}, \emph{14},
  273--284\relax
\mciteBstWouldAddEndPuncttrue
\mciteSetBstMidEndSepPunct{\mcitedefaultmidpunct}
{\mcitedefaultendpunct}{\mcitedefaultseppunct}\relax
\EndOfBibitem
\bibitem[Okoth \latin{et~al.}(2019)Okoth, Cavanna, Santiago-Cruz, and
  Chekhova]{Okoth2019}
Okoth,~C.; Cavanna,~A.; Santiago-Cruz,~T.; Chekhova,~M.~V. Microscale
  Generation of Entangled Photons without Momentum Conservation. \emph{Phys.
  Rev. Lett.} \textbf{2019}, \emph{123}, 263602\relax
\mciteBstWouldAddEndPuncttrue
\mciteSetBstMidEndSepPunct{\mcitedefaultmidpunct}
{\mcitedefaultendpunct}{\mcitedefaultseppunct}\relax
\EndOfBibitem
\bibitem[Okoth \latin{et~al.}(2020)Okoth, Kovlakov, B\"onsel, Cavanna, Straupe,
  Kulik, and Chekhova]{Okoth2020}
Okoth,~C.; Kovlakov,~E.; B\"onsel,~F.; Cavanna,~A.; Straupe,~S.; Kulik,~S.~P.;
  Chekhova,~M.~V. Idealized Einstein-Podolsky-Rosen states from
  non--phase-matched parametric down-conversion. \emph{Phys. Rev. A}
  \textbf{2020}, \emph{101}, 011801\relax
\mciteBstWouldAddEndPuncttrue
\mciteSetBstMidEndSepPunct{\mcitedefaultmidpunct}
{\mcitedefaultendpunct}{\mcitedefaultseppunct}\relax
\EndOfBibitem
\bibitem[Liu \latin{et~al.}(2016)Liu, Sinclair, Saravi, Keeler, Yang, Reno,
  Peake, Setzpfandt, Staude, Pertsch, and Brener]{Liu2016}
Liu,~S.; Sinclair,~M.~B.; Saravi,~S.; Keeler,~G.~A.; Yang,~Y.; Reno,~J.;
  Peake,~G.~M.; Setzpfandt,~F.; Staude,~I.; Pertsch,~T.; Brener,~I. Resonantly
  Enhanced Second-Harmonic Generation Using III--V Semiconductor All-Dielectric
  Metasurfaces. \emph{Nano Letters} \textbf{2016}, \emph{16}, 5426--5432\relax
\mciteBstWouldAddEndPuncttrue
\mciteSetBstMidEndSepPunct{\mcitedefaultmidpunct}
{\mcitedefaultendpunct}{\mcitedefaultseppunct}\relax
\EndOfBibitem
\bibitem[Vabishchevich \latin{et~al.}(2018)Vabishchevich, Liu, Sinclair,
  Keeler, Peake, and Brener]{Vabishchevich2018}
Vabishchevich,~P.~P.; Liu,~S.; Sinclair,~M.~B.; Keeler,~G.~A.; Peake,~G.~M.;
  Brener,~I. Enhanced Second-Harmonic Generation Using Broken Symmetry III--V
  Semiconductor Fano Metasurfaces. \emph{ACS Photonics} \textbf{2018},
  \emph{5}, 1685--1690\relax
\mciteBstWouldAddEndPuncttrue
\mciteSetBstMidEndSepPunct{\mcitedefaultmidpunct}
{\mcitedefaultendpunct}{\mcitedefaultseppunct}\relax
\EndOfBibitem
\bibitem[Liu \latin{et~al.}(2018)Liu, Vabishchevich, Vaskin, Reno, Keeler,
  Sinclair, Staude, and Brener]{Liu2018}
Liu,~S.; Vabishchevich,~P.~P.; Vaskin,~A.; Reno,~J.~L.; Keeler,~G.~A.;
  Sinclair,~M.~B.; Staude,~I.; Brener,~I. An all-dielectric metasurface as a
  broadband optical frequency mixer. \emph{Nature Communications}
  \textbf{2018}, \emph{9}, 2507\relax
\mciteBstWouldAddEndPuncttrue
\mciteSetBstMidEndSepPunct{\mcitedefaultmidpunct}
{\mcitedefaultendpunct}{\mcitedefaultseppunct}\relax
\EndOfBibitem
\bibitem[Koshelev \latin{et~al.}(2020)Koshelev, Kruk, Melik-Gaykazyan, Choi,
  Bogdanov, Park, and Kivshar]{Koshelev2020}
Koshelev,~K.; Kruk,~S.; Melik-Gaykazyan,~E.; Choi,~J.-H.; Bogdanov,~A.;
  Park,~H.-G.; Kivshar,~Y. Subwavelength dielectric resonators for nonlinear
  nanophotonics. \emph{Science} \textbf{2020}, \emph{367}, 288--292\relax
\mciteBstWouldAddEndPuncttrue
\mciteSetBstMidEndSepPunct{\mcitedefaultmidpunct}
{\mcitedefaultendpunct}{\mcitedefaultseppunct}\relax
\EndOfBibitem
\bibitem[Poddubny and Smirnova(2018)Poddubny, and Smirnova]{Poddubny2018}
Poddubny,~A.~N.; Smirnova,~D.~A. Nonlinear generation of quantum-entangled
  photons from high-Q states in dielectric nanoparticles. 2018;
  arXiv:1808.04811\relax
\mciteBstWouldAddEndPuncttrue
\mciteSetBstMidEndSepPunct{\mcitedefaultmidpunct}
{\mcitedefaultendpunct}{\mcitedefaultseppunct}\relax
\EndOfBibitem
\bibitem[Marino \latin{et~al.}(2019)Marino, Solntsev, Xu, Gili, Carletti,
  Poddubny, Rahmani, Smirnova, Chen, Lema\^{i}tre, Zhang, Zayats, Angelis, Leo,
  Sukhorukov, and Neshev]{Marino2019}
Marino,~G. \latin{et~al.}  Spontaneous photon-pair generation from a dielectric
  nanoantenna. \emph{Optica} \textbf{2019}, \emph{6}, 1416--1422\relax
\mciteBstWouldAddEndPuncttrue
\mciteSetBstMidEndSepPunct{\mcitedefaultmidpunct}
{\mcitedefaultendpunct}{\mcitedefaultseppunct}\relax
\EndOfBibitem
\bibitem[Ma \latin{et~al.}(2020)Ma, Ren, Wu, Cai, and Xu]{Ma2020}
Ma,~J.; Ren,~M.; Wu,~W.; Cai,~W.; Xu,~J. Resonantly tunable second harmonic
  generation from lithium niobate metasurfaces. 2020; arXiv:2002.06594\relax
\mciteBstWouldAddEndPuncttrue
\mciteSetBstMidEndSepPunct{\mcitedefaultmidpunct}
{\mcitedefaultendpunct}{\mcitedefaultseppunct}\relax
\EndOfBibitem
\bibitem[Solntsev \latin{et~al.}(2020)Solntsev, Agarwal, and
  Kivshar]{Solntsev2020}
Solntsev,~A.~S.; Agarwal,~G.~S.; Kivshar,~Y.~S. Metasurfaces for Quantum
  Photonics. 2020; arXiv:2007.14722\relax
\mciteBstWouldAddEndPuncttrue
\mciteSetBstMidEndSepPunct{\mcitedefaultmidpunct}
{\mcitedefaultendpunct}{\mcitedefaultseppunct}\relax
\EndOfBibitem
\bibitem[Koshelev \latin{et~al.}(2020)Koshelev, Bogdanov, and
  Kivshar]{Koshelev2020a}
Koshelev,~K.; Bogdanov,~A.; Kivshar,~Y. Engineering with bound states in the
  continuum. \emph{Opt. Photon. News} \textbf{2020}, \emph{31}, 38--45\relax
\mciteBstWouldAddEndPuncttrue
\mciteSetBstMidEndSepPunct{\mcitedefaultmidpunct}
{\mcitedefaultendpunct}{\mcitedefaultseppunct}\relax
\EndOfBibitem
\bibitem[Parry \latin{et~al.}(2019)Parry, Xing, Xu, Poddubny, Neshev, and
  Sukhorukov]{Parry2019}
Parry,~M.; Xing,~Y.; Xu,~L.; Poddubny,~A.; Neshev,~D.~N.; Sukhorukov,~A.~A.
  {Photon-pair generation via bound states in the continuum in nonlinear
  metasurfaces}. SPIE Micro + Nano Materials, Devices, and Applications 2019.
  2019; pp 20 -- 21\relax
\mciteBstWouldAddEndPuncttrue
\mciteSetBstMidEndSepPunct{\mcitedefaultmidpunct}
{\mcitedefaultendpunct}{\mcitedefaultseppunct}\relax
\EndOfBibitem
\bibitem[Wang \latin{et~al.}(2019)Wang, Li, and Zhang]{Wang2019}
Wang,~T.; Li,~Z.; Zhang,~X. Improved generation of correlated photon pairs from
  monolayer WS2 based on bound states in the continuum. \emph{Photon. Res.}
  \textbf{2019}, \emph{7}, 341--350\relax
\mciteBstWouldAddEndPuncttrue
\mciteSetBstMidEndSepPunct{\mcitedefaultmidpunct}
{\mcitedefaultendpunct}{\mcitedefaultseppunct}\relax
\EndOfBibitem
\bibitem[Dinparasti~Saleh \latin{et~al.}(2018)Dinparasti~Saleh, Vezzoli,
  Caspani, Branny, Kumar, Gerardot, and Faccio]{Faccio2018}
Dinparasti~Saleh,~H.; Vezzoli,~S.; Caspani,~L.; Branny,~A.; Kumar,~S.;
  Gerardot,~B.~D.; Faccio,~D. Towards spontaneous parametric down conversion
  from monolayer MoS2. \emph{Scientific Reports} \textbf{2018}, \emph{8},
  3862\relax
\mciteBstWouldAddEndPuncttrue
\mciteSetBstMidEndSepPunct{\mcitedefaultmidpunct}
{\mcitedefaultendpunct}{\mcitedefaultseppunct}\relax
\EndOfBibitem
\bibitem[Lee \latin{et~al.}(2017)Lee, Tian, Yang, Mustonen, Martinez, Dai,
  Kauppinen, Malowicki, Kumar, and Sun]{Lee2017}
Lee,~K.~F.; Tian,~Y.; Yang,~H.; Mustonen,~K.; Martinez,~A.; Dai,~Q.;
  Kauppinen,~E.~I.; Malowicki,~J.; Kumar,~P.; Sun,~Z. Photon-Pair Generation
  with a 100 nm Thick Carbon Nanotube Film. \emph{Advanced Materials}
  \textbf{2017}, \emph{29}, 1605978\relax
\mciteBstWouldAddEndPuncttrue
\mciteSetBstMidEndSepPunct{\mcitedefaultmidpunct}
{\mcitedefaultendpunct}{\mcitedefaultseppunct}\relax
\EndOfBibitem
\bibitem[Cambiasso \latin{et~al.}(2017)Cambiasso, Grinblat, Li, Rakovich,
  Cort{\'e}s, and Maier]{Cambiasso2017}
Cambiasso,~J.; Grinblat,~G.; Li,~Y.; Rakovich,~A.; Cort{\'e}s,~E.; Maier,~S.~A.
  Bridging the Gap between Dielectric Nanophotonics and the Visible Regime with
  Effectively Lossless Gallium Phosphide Antennas. \emph{Nano Letters}
  \textbf{2017}, \emph{17}, 1219--1225\relax
\mciteBstWouldAddEndPuncttrue
\mciteSetBstMidEndSepPunct{\mcitedefaultmidpunct}
{\mcitedefaultendpunct}{\mcitedefaultseppunct}\relax
\EndOfBibitem
\bibitem[Shcherbakov \latin{et~al.}(2020)Shcherbakov, Zhang, Tripepi,
  Sartorello, Talisa, AlShafey, Fan, Twardowski, Krivitsky, Kuznetsov,
  Chowdhury, and Shvets]{Shcherbakov2020}
Shcherbakov,~M.~R.; Zhang,~H.; Tripepi,~M.; Sartorello,~G.; Talisa,~N.;
  AlShafey,~A.; Fan,~Z.; Twardowski,~J.; Krivitsky,~L.~A.; Kuznetsov,~A.~I.;
  Chowdhury,~E.; Shvets,~G. Generation of even and odd high harmonics in
  resonant metasurfaces using single and multiple ultra-intense laser pulses.
  2020; arXiv:2008.03619\relax
\mciteBstWouldAddEndPuncttrue
\mciteSetBstMidEndSepPunct{\mcitedefaultmidpunct}
{\mcitedefaultendpunct}{\mcitedefaultseppunct}\relax
\EndOfBibitem
\bibitem[Fedotova \latin{et~al.}(2019)Fedotova, Younesi, Sautter, Steinert,
  Geiss, Pertsch, Staude, and Setzpfandt]{Fedotova2019}
Fedotova,~A.; Younesi,~M.; Sautter,~J.; Steinert,~M.; Geiss,~R.; Pertsch,~T.;
  Staude,~I.; Setzpfandt,~F. Second-Harmonic Generation in Lithium Niobate
  Metasurfaces. 2019 Conference on Lasers and Electro-Optics Europe and
  European Quantum Electronics Conference. 2019\relax
\mciteBstWouldAddEndPuncttrue
\mciteSetBstMidEndSepPunct{\mcitedefaultmidpunct}
{\mcitedefaultendpunct}{\mcitedefaultseppunct}\relax
\EndOfBibitem
\bibitem[Loudon(2000)]{Loudon2000}
Loudon,~R. \emph{The Quantum Theory of Light}; OUP Oxford, 2000\relax
\mciteBstWouldAddEndPuncttrue
\mciteSetBstMidEndSepPunct{\mcitedefaultmidpunct}
{\mcitedefaultendpunct}{\mcitedefaultseppunct}\relax
\EndOfBibitem
\bibitem[Anthur \latin{et~al.}(2020)Anthur, Zhang, Akimov, Ong, Kalashnikov,
  Kuznetsov, and Krivitsky]{Anthur2020}
Anthur,~A.~P.; Zhang,~H.; Akimov,~Y.; Ong,~J.; Kalashnikov,~D.;
  Kuznetsov,~A.~I.; Krivitsky,~L. Demonstration of second harmonic generationin
  gallium phosphide nano-waveguides. 2020; arXiv:2001.06142\relax
\mciteBstWouldAddEndPuncttrue
\mciteSetBstMidEndSepPunct{\mcitedefaultmidpunct}
{\mcitedefaultendpunct}{\mcitedefaultseppunct}\relax
\EndOfBibitem
\bibitem[Valencia \latin{et~al.}(2002)Valencia, Chekhova, Trifonov, and
  Shih]{Valencia2002}
Valencia,~A.; Chekhova,~M.~V.; Trifonov,~A.; Shih,~Y. Entangled Two-Photon Wave
  Packet in a Dispersive Medium. \emph{Phys. Rev. Lett.} \textbf{2002},
  \emph{88}, 183601\relax
\mciteBstWouldAddEndPuncttrue
\mciteSetBstMidEndSepPunct{\mcitedefaultmidpunct}
{\mcitedefaultendpunct}{\mcitedefaultseppunct}\relax
\EndOfBibitem
\bibitem[Kitaeva \latin{et~al.}(1982)Kitaeva, Klyshko, and Taubin]{Kitaeva1982}
Kitaeva,~G.~K.; Klyshko,~D.~N.; Taubin,~I.~V. Theory of parametric scattering
  and method of absolute measurement of the brightness of light. \emph{Soviet
  Journal of Quantum Electronics} \textbf{1982}, \emph{12}, 333--338\relax
\mciteBstWouldAddEndPuncttrue
\mciteSetBstMidEndSepPunct{\mcitedefaultmidpunct}
{\mcitedefaultendpunct}{\mcitedefaultseppunct}\relax
\EndOfBibitem
\bibitem[Li \latin{et~al.}(2020)Li, Liu, Ren, Wang, Su, Chen, Chu, Kuo, Liu,
  Zang, Guo, Zhang, Wang, Zhu, and Tsai]{Li2020}
Li,~L.; Liu,~Z.; Ren,~X.; Wang,~S.; Su,~V.-C.; Chen,~M.-K.; Chu,~C.~H.;
  Kuo,~H.~Y.; Liu,~B.; Zang,~W.; Guo,~G.; Zhang,~L.; Wang,~Z.; Zhu,~S.;
  Tsai,~D.~P. Metalens-array{\textendash}based high-dimensional and multiphoton
  quantum source. \emph{Science} \textbf{2020}, \emph{368}, 1487--1490\relax
\mciteBstWouldAddEndPuncttrue
\mciteSetBstMidEndSepPunct{\mcitedefaultmidpunct}
{\mcitedefaultendpunct}{\mcitedefaultseppunct}\relax
\EndOfBibitem
\end{mcitethebibliography}

\end{document}